\documentclass[twocolumn,showpacs,preprintnumbers,amsmath,amssymb]{revtex4-1}

\usepackage{graphicx}
\usepackage{dcolumn}
\usepackage{bm}
\usepackage{color}

\begin{document}

\preprint{}

\title{Glueball scattering cross section in lattice $SU(2)$ Yang-Mills theory}

\author{Nodoka Yamanaka$^{1,2}$}
\email{ynodoka@yukawa.kyoto-u.ac.jp}
\author{Hideaki Iida$^3$}
\author{Atsushi Nakamura$^{4,5,6}$}
\author{Masayuki Wakayama$^{7,8,9,5}$}
\affiliation{$^1$Yukawa Institute for Theoretical Physics, Kyoto University, 
  Kitashirakawa-Oiwake, Kyoto 606-8502, Japan}
\affiliation{$^2$Amherst Center for Fundamental Interactions, Department of Physics, University of Massachusetts Amherst, MA 01003, USA}
\affiliation{$^3$Department of Physics, The University of Tokyo, 7-3-1 Hongo, Bunkyo-ku, Tokyo 113-0033, Japan}
\affiliation{$^4$Pacific Quantum Center, Far Eastern Federal University, Sukhanova 8, Vladivostok, 690950, Russia}
\affiliation{$^5$Research Center for Nuclear Physics, Osaka University, Ibaraki, Osaka 567-0047, Japan}
\affiliation{$^6$Theoretical Research Division, Nishina Center, RIKEN, 
  Wako, Saitama 351-0198, Japan}
\affiliation{$^7$School of Science and Engineering, Kokushikan University, Tokyo 154-8515, Japan}
\affiliation{$^8$Center for Extreme Nuclear Matters (CENuM), Korea University, Seoul 02841, Republic of Korea}
\affiliation{$^9$Department of Physics, Pukyong National University (PKNU), Busan 48513, Republic of Korea}

\date{\today}

\begin{abstract}
We calculate the scattering cross section between two $0^{++}$ glueballs in $SU(2)$ Yang-Mills theory on lattice at $\beta = 2.1, 2.2, 2.3, 2.4$, and 2.5 using the indirect (HAL QCD) method.
We employ the cluster-decomposition error reduction technique and use all space-time symmetries to improve the signal.
In the use of the HAL QCD method, the centrifugal force was subtracted to remove the systematic effect due to nonzero angular momenta of lattice discretization.
From the extracted interglueball potential we determine the low energy glueball effective theory by matching with the one-glueball exchange process.
We then calculate the scattering phase shift, and derive the relation between the interglueball cross section and the scale parameter $\Lambda$ as $\sigma_{\phi \phi} = (2 - 51) \Lambda^{-2}$ (stat.+sys.).
From the observational constraints of galactic collisions, we obtain the lower bound of the scale parameter, as $\Lambda > 60$ MeV.
We also discuss the naturalness of the Yang-Mills theory as the theory explaining dark matter.
\end{abstract}

\pacs{11.15.-q,12.39.Mk,95.35.+d,98.80.-k}

\maketitle

\section{\label{sec:intro}Introduction}

The study of dark matter (DM) \cite{Bertone:2004pz,Munshi:2006fn,Arcadi:2017kky,Battaglieri:2017aum} is one of the most fundamental subjects of physics.
Its presence is providing us the most consistent explanation for many astrophysical and cosmological phenomena and problems such as the large scale structure formation \cite{Blumenthal:1984bp,Navarro:1995iw,Navarro:1996gj,Hu:1998kj,Yoshida:2000uw,Jing:2002np,Evans:2008mp,Oguri:2010ik,Ishiyama:2011af,Oguri:2011dt,Frenk:2012ph,Okabe:2013efa,Ishiyama:2014uoa,Umetsu:2015baa,Niikura:2015uda,Clampitt:2015wea,vanUitert:2016guv,Shin:2017rch,Umetsu:2018ypz}.
While the DM is less likely to be mainly composed of astrophysical objects \cite{Tisserand:2006zx,Ricotti:2007au,Carr:2009jm,Barnacka:2012bm,Capela:2013yf,Griest:2013esa,Ali-Haimoud:2016mbv,Niikura:2017zjd,Oguri:2017ock,Zumalacarregui:2017qqd,Sasaki:2018dmp}, there are no particle physics candidates in the standard model (SM), and many models beyond it are under investigation \cite{Bertone:2004pz,Arcadi:2017kky,Battaglieri:2017aum,Dine:1982ah,Preskill:1982cy,Barr:1990ca,Barr:1991qn,Kaplan:1991ah,Dodelson:1993je,Jungman:1995df,Shi:1998km,Moroi:1999zb,Feng:2000gh,Cheng:2002ej,Servant:2002aq,Narain:2006kx,LopezHonorez:2006gr,Boehmer:2007um,Pospelov:2007mp,MarchRussell:2008yu,Pospelov:2008zw,ArkaniHamed:2008qp,Hambye:2008bq,Zurek:2008qg,Cheung:2009qd,Hosotani:2009jk,Hambye:2009pw,Morrissey:2009ur,Feng:2009mn,Goodsell:2009xc,DEramo:2010keq,Feng:2010gw,Shelton:2010ta,Davoudiasl:2010am,Buckley:2010ui,Falkowski:2011xh,Lin:2011gj,Lebedev:2011iq,LopezHonorez:2012kv,Hooper:2012cw,Carone:2013wla,Adhikari:2016bei,Hui:2016ltb,Dienes:2016vei,Hambye:2016qkf,Roszkowski:2017nbc,Plehn:2017fdg,Ho:2017fte,Chu:2017msm,Knapen:2017xzo,Falkowski:2018dsl,Carney:2019pza,Mohamadnejad:2019vzg,Keus:2019szx}.
The ``WIMP miracle'' \cite{Griest:1989wd,Griest:1990kh,Jungman:1995df,McDonald:2001vt,Chu:2011be,Blum:2014dca,Berlin:2016gtr,DAgnolo:2018wcn,Batell:2018fqo,Lague:2019yvs}, i.e. the suggestive coincidence of the weak coupling and the DM density that would have thermally been generated with the latter as the interaction between ordinary matter and DM, motivated us to experimentally test their direct scattering  \cite{Servant:2002hb,Belanger:2008sj,Ahmed:2009zw,Djouadi:2012zc,Aprile:2012nq,Agnese:2013rvf,Bernabei:2013xsa,Jia:2017iyc,Bertuzzo:2017lwt,Cui:2017nnn,Ren:2018gyx,Crisler:2018gci,Xia:2018qgs,Aguilar-Arevalo:2019wdi,Liu:2019kzq}, the DM decay to visible cosmic rays \cite{Strong:1998pw,Baltz:1998xv,Bergstrom:2004cy,Hisano:2004ds,ArkaniHamed:2008qn,Cirelli:2008pk,Ibe:2013nka,Adriani:2008zr,Pospelov:2008jd,Chen:2008qs,FermiLAT:2011ab,Adriani:2013uda,Aartsen:2013jdh,Accardo:2014lma,Aguilar:2014fea,TheFermi-LAT:2015kwa,Aartsen:2016ngq,Aguilar:2016kjl,Aartsen:2017nbu,Ambrosi:2017wek,Aguilar:2019owu,Murase:2019tjj,DeRujula:2019wbk}, and DM productions at collider experiments \cite{Baer:2005bu,Falkowski:2010cm,Beltran:2010ww,Goodman:2010yf,Bai:2010hh,Goodman:2010ku,Fox:2011pm,Djouadi:2011aa,Chatrchyan:2012tea,Djouadi:2012zc,Chatrchyan:2012me,Carpenter:2013xra,Lees:2014xha,Khachatryan:2014rra,Aad:2015zva,Chala:2015ama,Abdallah:2015ter,Aad:2015pla,Abercrombie:2015wmb,Khachatryan:2016mbu,Sirunyan:2016iap,Khachatryan:2016whc,Kahlhoefer:2017dnp,Aaboud:2017phn,Beauchesne:2017yhh,Sirunyan:2018xlo,Belyaev:2018pqr,Abe:2018bpo}, but no positive results have been reported so far.

In this context, an additional Yang-Mills theory (YMT) which does not or very weakly interacts with the SM particles is an attractive candidate, since the lightest particle of the spectrum is a glueball \cite{Jaffe:1975fd,Coleman:1977hd,Robson:1977pm,Ishikawa:1978vc,Novikov:1979va,Suura:1980ex,Schechter:1980ak,Salomone:1980sp,Donoghue:1980hw,Carlson:1980bg,Carlson:1980kh,Shifman:1980iq,Cornwall:1981zr,Hansson:1982dv,Migdal:1982jp,Pascual:1982bv,Mandula:1982us,Cornwall:1982zn,Ward:1984in,Ellis:1984jv,Gomm:1984zq,Jaffe:1985qp,Bordes:1989kc,Gensini:1989ch,Bagan:1989vm,Bagan:1990sy,Demeterfi:1993rs,Birse:1994ar,Schafer:1994fd,Csaki:1998qr,Minahan:1998tm,deMelloKoch:1998vqw,Zyskin:1998tg,Ooguri:1998hq,Constable:1999gb,Kaidalov:1999yd,Kaidalov:1999de,Bali:2000gf,Brower:2000rp,Caceres:2000qe,Hou:2001ig,Hou:2002jv,Dijkgraaf:2002xd,Ita:2002kx,BoschiFilho:2002vd,Forkel:2003mk,Szczepaniak:2003mr,Aganagic:2003xq,BoschiFilho:2002ta,Brau:2004xw,Meyer:2004gx,Vento:2004xx,Leigh:2006vg,Kondo:2006sa,Colangelo:2007pt,Klempt:2007cp,Forkel:2007ru,Mathieu:2008bf,Boulanger:2008aj,Mathieu:2008me,Dudal:2008tg,Kharzeev:2008br,Dudal:2009zh,Colangelo:2009ra,Dudal:2010cd,Vandersickel:2011zc,Ochs:2013gi,Li:2013oda,Dudal:2013wja,Brambilla:2014jmp,FolcoCapossoli:2016ejd,Elander:2017hyr,Hong:2017suj,Ballon-Bayona:2017sxa,Rinaldi:2018yhf,Rinaldi:2020ssz}, fulfilling the conditions required for being DM \cite{Carlson:1992fn,Faraggi:2000pv,Juknevich:2009ji,Juknevich:2009gg,Feng:2011ik,Boddy:2014yra,Soni:2016gzf,Kribs:2016cew,Forestell:2016qhc,Halverson:2016nfq,Soni:2016yes,Acharya:2017szw,Soni:2017nlm,daRocha:2017cxu,Forestell:2017wov,Brower:2019oor,Kang:2019izi,Hertzberg:2019bvt,Buttazzo:2019iwr}.
This ``dark'' YMT may naturally be generated in grand unification frameworks \cite{Ellis:1990sr,Faraggi:1992fa,Carlson:1992fn,Leontaris:1995sf,Dienes:1996du,Kakushadze:1996tm,Kakushadze:1996jm,Kakushadze:1996iw,Kakushadze:1997ne,Benakli:1998ut,Faraggi:2000pv,Halverson:2016nfq,DelleRose:2017vvz,McGuigan:2019gdb,Ibe:2019ena}, and it is possible to experimentally detect them by observing the gravitational wave background \cite{Kamionkowski:1993fg,Maggiore:1999vm,Kosowsky:2001xp,Dolgov:2002ra,Grojean:2006bp,Caprini:2007xq,Huber:2008hg,Caprini:2009fx,Caprini:2009yp,Hindmarsh:2013xza,Hindmarsh:2015qta,Schwaller:2015tja,Caprini:2015zlo,Soni:2016yes,Fairbairn:2019xog,Hindmarsh:2019phv,Alanne:2019bsm} left from the first order de/confinement phase transition (YMTs have first order de/confinement phase transition for color number $N_c \ge 3$ \cite{Svetitsky:1982gs,Wingate:2000bb,Gavai:2002td,Lucini:2002ku,Lucini:2003zr,Lucini:2005vg,Lucini:2012wq,Lucini:2012gg}).

The glueballs, like the other hadrons, are generated by the nonperturbative physics of nonabelian gauge theory, so the lattice gauge theory simulation is required to quantify their dynamics \cite{Berg:1980gz,Nauenberg:1981kv,Munster:1981es,Engels:1981wu,Ishikawa:1982tb,Mutter:1982cx,Brower:1982fy,Berg:1981zb,Berg:1982mu,Smit:1982fx,Falcioni:1982ja,Seo:1982jh,Ishikawa:1983xg,Azcoiti:1983ab,deForcrand:1984eeq,Albanese:1987ds,Teper:1987wt,Teper:1987ws,deForcrand:1991kc,Bali:1993fb,Sexton:1995kd,Teper:1998kw,Vaccarino:1999ku,Morningstar:1999rf,Bali:2000vr,Lucini:2001ej,Ishii:2001zq,Ishii:2002ww,Meyer:2003wx,Lucini:2004my,Chen:2005mg,Loan:2006gm,Orland:2007rj,Lucini:2010nv,Gregory:2012hu,Lucini:2012gg,Chowdhury:2014kfa,Teper:2018qvw}.
The glueballs should also exist in the quantum chromodynamics of the SM, and extensive experimental search is ongoing \cite{Crede:2008vw,Binon:1984ip,Akesson:1985rn,Barberis:1996iq,Barate:1999ze,Acciarri:2000ex,Abbiendi:2003ri,Carman:2005ps,Ablikim:2006db,Chekanov:2008ad,Uehara:2013mbo}, but the mixing with other hadrons \cite{Rosner:1981sc,Schnitzer:1981zg,Lipkin:1983kd,Lanik:1984fc,Wong:1986gm,Amsler:1995td,Amsler:1995tu,Anisovich:1996zj,Anisovich:1997ye,Minkowski:1998mf,StrohmeierPresicek:1999yv,Ellis:1999kv,Close:2000yk,Close:2001ga,Fariborz:2003uj,Giacosa:2004ug,Cheng:2006hu,Fariborz:2006xq,Mennessier:2008kk,Albaladejo:2008qa,Nussinov:2009tq,Chen:2009zzs,RuizArriola:2010fj,Janowski:2011gt,Janowski:2014ppa,Wakayama:2014gpa,Cheng:2015iaa,Vento:2015yja,Pelaez:2015qba,Zou:2019tpo,Soni:2019xko} is complicating the analysis of their production and decay processes \cite{Coyne:1980zd,Chanowitz:1983sd,Cornwall:1983zb,Gershtein:1983kc,Cornwall:1984pa,Gensini:1989ch,Cakir:1994jf,Close:1997pj,Giacosa:2005qr,Chanowitz:2005du,Chao:2005si,Zhao:2005ip,Giacosa:2005zt,He:2006qk,Chao:2007sk,Chanowitz:2007ma,Chen:2007uq,Hashimoto:2007ze,Jin:2008zza,Cheng:2010ry,Cohen:2014vta,Brunner:2015yha,Frere:2015xxa,Parganlija:2016yxq,Brodsky:2018snc,Llanes-Estrada:2018omz,Wan:2019fuk,Khoze:2019tcn}.
In the case of the DM, the glueballs of the YMT are stable.
The absence of quarks is indeed an important advantage, since the hierarchy problem will become almost irrelevant.

Here we challenge the quantification of the interglueball cross section.
The DM self-interaction (scattering) may affect the structure of the galactic halos and their collisions \cite{deLaix:1995vi,Spergel:1999mh,Kamionkowski:1999vp,Ostriker:1999ee,Mohapatra:2000qx,Moore:2000fp,Hannestad:2000gt,Kochanek:2000pi,Binney:2000zt,Hogan:2000bv,Peebles:2000yy,vandenBosch:2000rza,Riotto:2000kh,Oh:2010mc,Vogelsberger:2012ku,Rocha:2012jg,Zavala:2012us,Peter:2012jh,Tulin:2013teo,Kahlhoefer:2013dca,Hochberg:2014dra,Brook:2015qja,Bernal:2015ova,Cyr-Racine:2015ihg,Kamada:2016euw,Tulin:2017ara,Bernal:2017mqb,Chu:2018fzy,Zavala:2019gpq,Chu:2019awd}.
The DM scattering cross section is actually constrained by observations \cite{Markevitch:2003at,Randall:2007ph,Bradac:2008eu,Merten:2011wj,Harvey:2015hha}.
We expect the lattice calculation of the interglueball scattering to yield quantitative relation between the cross section and the unknown scale parameter $\Lambda$ of the dark YMT, which will be bounded by the observational data.
In this work, we use the HAL QCD method \cite{Ishii:2006ec,Aoki:2009ji,Inoue:2010hs,Inoue:2010es,Ikeda:2011bs,Kawanai:2011xb,Doi:2011gq,HALQCD:2012aa,Murano:2013xxa,Ikeda:2013vwa,Sasaki:2015ifa,Ikeda:2016zwx,Gongyo:2017fjb,Miyamoto:2017tjs,Iritani:2018sra,Akahoshi:2019klc,Miyamoto:2019jjc}, which is quite successful in the determination of the interhadron potential on lattice, to quantify the glueball scattering.

This paper gives the complete and detailed discussion of the letter \cite{Yamanaka:2019aeq}, in which the DM cross section within the YMT was quantitatively derived for the first time on lattice.
In the next section, we will explain the naturalness of the YMT as the candidate of DM model beyond the SM.
We then describe in Section \ref{sec:formalism} the setup of lattice simulation and the calculation of the interglueball potential obtained by employing the HAL QCD method.
In Section \ref{sec:results}, we show the result of our calculation of the potential in $SU(2)$ lattice YMT, determine the glueball effective field theory by matching the one-glueball exchange process with our lattice data, derive the glueball cross section, and constrain the scale parameter from observation.
The final section is devoted to the summary.

\section{\label{sec:naturalness}Naturalness of dark Yang-Mills theory}

In this section, we shall show that the glueball DM is very natural among composite DM scenarios \cite{Nussinov:1985xr,Chivukula:1989qb,Gudnason:2006yj,Khlopov:2008ty,Ryttov:2008xe,Foadi:2008qv,Shepherd:2009sa,Alves:2009nf,Mardon:2009gw,Hambye:2009fg,Kaplan:2009de,Kribs:2009fy,Alves:2010dd,Bai:2010qg,Lewis:2011zb,Cline:2012is,Frigerio:2012uc,CyrRacine:2012fz,Buckley:2012ky,Appelquist:2013ms,Braaten:2013tza,Bai:2013xga,Bhattacharya:2013kma,Hietanen:2013fya,Cyr-Racine:2013fsa,Laha:2013gva,Cline:2013zca,Boddy:2014yra,Appelquist:2014jch,Krnjaic:2014xza,Detmold:2014qqa,Detmold:2014kba,Boddy:2014qxa,Foot:2014uba,Hochberg:2014kqa,Hardy:2014mqa,Foot:2014osa,Lonsdale:2014yua,Bian:2014cja,Bernal:2015bla,Appelquist:2015zfa,Appelquist:2015yfa,Antipin:2015xia,Carmona:2015haa,Lee:2015gsa,Hardy:2015boa,Laha:2015yoa,Buen-Abad:2015ova,Gross:2015cwa,Bernal:2015xba,Soni:2016gzf,Foot:2016wvj,Pappadopulo:2016pkp,Arthur:2016dir,Kribs:2016cew,Forestell:2016qhc,Bruggisser:2016ixa,Bruggisser:2016nzw,Farina:2016llk,Kopp:2016yji,Halverson:2016nfq,Ko:2016fcd,Lonsdale:2017mzg,Batell:2017kho,Berryman:2017twh,Choi:2017zww,Mitridate:2017oky,Davoudiasl:2017zws,Alexander:2018fjp,Braaten:2018xuw,Oncala:2018bvl,Francis:2018xjd,Beauchesne:2018myj,Kribs:2018oad,Kribs:2018ilo,Bai:2018dxf,Contino:2018crt,Ibe:2018tex,Redi:2018muu,Elahi:2019jeo,Terning:2019hgj,Brower:2019oor,Smirnov:2019ngs,Dondi:2019olm,Ibe:2019ena,Bernreuther:2019pfb,Ibe:2019yra,Mahbubani:2019pij,Hertzberg:2019bvt,Bennett:2019jzz} and dark gauge theories.
The assumption of the existence of dark gauge sectors is indeed a reasonable answer to the problem of the ad hoc gauge group of the SM, since they may naturally arise in many contexts such as in string or grand unification theories \cite{Ellis:1990sr,Faraggi:1992fa,Carlson:1992fn,Leontaris:1995sf,Dienes:1996du,Kakushadze:1996tm,Kakushadze:1996jm,Kakushadze:1996iw,Kakushadze:1997ne,Benakli:1998ut,Faraggi:2000pv,Halverson:2016nfq,DelleRose:2017vvz,McGuigan:2019gdb,Ibe:2019ena}.
Here we do not discuss their origin, but rather the constraint on the dark gauge sector that could be imposed assuming the naturalness.

The DM model has to be conceived respecting many phenomenological constraints such as the electric  neutrality, the nonrelativisticity, the consistency with the bigbang nucleosynthesis, experimental data from direct and indirect detections, etc.
The simplest possibility is the massive dark photon.
To give it a finite mass, the Higgs potential has to be present with a scalar quadratic term.
The latter is known to introduce severe hierarchical problem due to the quadratic divergence, unless the Higgs potential is just an effective one below some energy scale close to the dark photon mass.
The same reasoning applies for nonabelian gauge theories with scalar field(s), whether it (they) induces the Higgs mechanism or not.

The second class of theory is the gauge theory without spontaneous breakdown by the Higgs mechanism.
In this case, we can first conceive an asymptotically free gauge theory.
However, we have to note that the fermions must be massless (or very light compared to the current temperature of the Universe), which will conflict with the nonrelativistic property and the bigbang nucleosynthesis.
In this class of gauge theory, colored particles should therefore be confined, and the mass of hadrons is generated dynamically.
The chiral fermions are not favored for several reasons as shown below.
If the fermions are chiral i.e. massless, the general case will be the spontaneous break down of chiral symmetry.
Massless Nambu-Goldstone modes are then generated, which are also not allowed phenomenologically, due to the constraint on the number of relativistic particles at the bigbang nucleosynthesis.
We may equally think of a case where the chiral symmetry is not spontaneously broken, requiring the generation of massless composite fermions with the same global symmetry as the original theory, as required by the 't Hooft anomaly matching.
This scenario is again forbidden for the same reason as the previous case.
If the elementary fermions become massive due to other spontaneous breakdown of the chiral symmetry through the Yukawa interaction (like the SM), this means that there is at least one additional scalar field which forms the Higgs potential, thus generating again the hierarchical problem.
We finally arrive at the conclusion that the most natural scenarios are nonabelian gauge theories with vectorlike fermions.

In QCD-like theories, the vectorlike fermions may have arbitrary masses.
If the vectorlike fermions are lighter than the confining scale ($\Lambda_{N_c}$) and have gauge charges of the SM sector, the dark baryon number asymmetry may be generated through the sphaleron process.
On the contrary, if the masses of vectorlike fermions are heavier than $\Lambda_{N_c}$, the glueball of this gauge sector will become the DM.
If we wish to explain the DM within the nonabelian gauge theory with vectorlike fermions, we therefore have two choices, the DM composed of baryons made of vectorlike fermions, or lightest glueballs for which the YMT is the relevant theory.
Needless to say, we may also conceive a pure YMT without any other fields
\begin{equation}
{\cal L}_{\rm YM}
=
-\frac{1}{4} \sum_{a=1}^{N_c^2-1} F^{\mu \nu}_a F_{\mu \nu}^a
,
\end{equation}
where $F^{\mu \nu}_a$ is the dark gluon field strength.
In this case, the number of input parameters is minimal, so that the YMT is the most natural theory explaining the DM.
Another feature which has to be emphasized is that, in grand unification scenarios, $\Lambda_{N_c}$ is controlled by the integer number of colors $N_c$ which runs the coupling logarithmically over the energy scale, and it may generate a variety of energy scales.

\section{\label{sec:formalism}Simulation and formalism}

\subsection{Simulation setup}

Let us now present the detail of the simulation of YMT on lattice.
In this work, we simulate it with five lattice spacings corresponding to $\beta = 2.1, 2.2, 2.3, 2.4$, and 2.5.
The simulation parameters are given in Table \ref{table:simulationparameters}.

\begin{table}[htb]
\caption{
Simulation parameters of $SU(2)$ YMT.
``Thermalization'' denotes the number of thermalization sweeps used in the pseudo-heat method, and ``Separation'' is the interval of sweeps taken between each data taking.
}
\begin{ruledtabular}
\begin{tabular}{l|cccc}
$\beta$ & Volume & Configurations & Thermalization & Separation \\ 
\hline
2.1 & $10^3 \times 12$ & 1000000 & 5000 & 150 \\
\hline
2.2 & $12^4$ & 9999990 & 5000 & 150 \\
\hline
2.3 & $14^3 \times 16$ & 4100000 & 8000 & 240 \\
\hline
2.4 & $16^3 \times 24$ & 2030000 & 5000 & 150 \\
\hline
2.5 & $20^3 \times 24$ & 520000 & 12000 & 600 \\
\end{tabular}
\end{ruledtabular}
\label{table:simulationparameters}
\end{table}

The relation between the string tension $\sigma$ and $\Lambda$, the scale parameter of $SU(2)$ YMT, was calculated for the general $N_c$ \cite{Allton:2008ty,Teper:2009uf}, giving
\begin{eqnarray}
\frac{\Lambda}{\sqrt{\sigma}}
=
0.503(2)(40)+ \frac{0.33(3)(3)}{N_c^2}
.
\label{eq:stringtensionNc}
\end{eqnarray}
By using the result of the calculation of the string tension of Ref. \cite{Teper:1998kw} (see Table \ref{table:stringtension}), it is possible to express the lattice spacings $a$ in the unit of $\Lambda$.
We note that $\Lambda$ is an unknown parameter, since the property of the DM particle is totally unknown.
The aim of our work is precisely to calculate the low energy constants of the low energy glueball effective Lagrangian and the interglueball scattering cross section in the unit of $\Lambda$, so that the constraint from observational data will give a bound on it.

\begin{table}[htb]
\caption{
The string tensions of Ref. \cite{Teper:1998kw} and lattice spacings derived from Eq. (\ref{eq:stringtensionNc}) for several $\beta$.
}
\begin{ruledtabular}
\begin{tabular}{l|cc}
$\beta$ & $a \sqrt{\sigma}$ & $a [\Lambda^{-1}]$ \\ 
\hline
2.1 & 0.608(16) & 0.356(27) \\
\hline
2.2 & 0.467(10) & 0.273(20) \\
\hline
2.3 & 0.3687(22) & 0.216(15) \\
\hline
2.4 & 0.2660(21) & 0.156(11) \\
\hline
2.5 & 0.1881(28) & 0.110(8) \\
\end{tabular}
\end{ruledtabular}
\label{table:stringtension}
\end{table}

Let us now define the $0^{++}$ glueball operator:
\begin{eqnarray}
\phi (t, \vec{x}) 
&=&
{\rm Re} [
P_{12} (t, \vec{x}) 
+P_{12} (t, \vec{x}+a\vec{e}_3) 
+P_{23} (t, \vec{x}) 
\nonumber\\
&& 
+P_{23} (t, \vec{x}+a\vec{e}_1) 
+P_{31} (t, \vec{x}) 
+P_{31} (t, \vec{x}+a\vec{e}_2) 
]
.
\nonumber\\
\label{eq:glueballop}
\end{eqnarray}
Here $P_{ij}$ are the plaquette operators in the $i-j$ direction, with $\vec{e}_{1,2,3}$ the unit vector.
The $0^{++}$ glueball has the same quantum number as the vacuum, so it has an expectation value which corresponds to a divergence in the continuum limit.
To extract physical information, we have to subtract it from the glueball field operator $\phi$.
The glueball correlators are then expressed in terms of $\tilde \phi (t, \vec{x}) \equiv \phi (t, \vec{x}) - \langle \phi (t, \vec{x}) \rangle$.

\begin{figure}[htb]
\includegraphics[width=8.5cm]{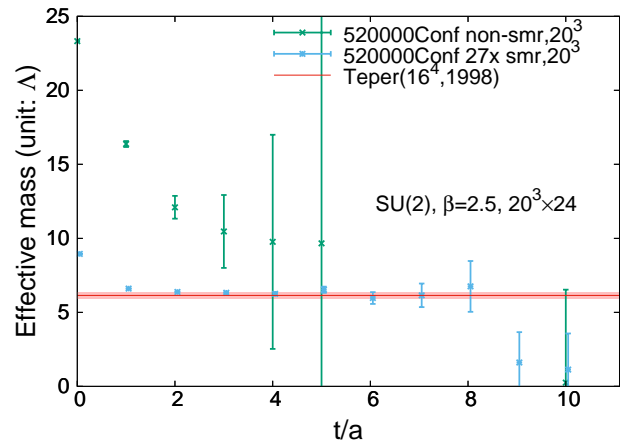}
\caption{\label{fig:effmass}
Glueball effective mass calculated with the standard glueball operator (\ref{eq:glueballop}) (green points) and with APE smearing (blue points) at $\beta=2.5$.
The result of the previous work \cite{Teper:1998kw} (red line, with the uncertainty band) is also shown for comparison.
}
\end{figure}

In order to improve the glueball operator, we also apply the APE smearing \cite{Albanese:1987ds,Ishii:2001zq,Ishii:2002ww}.
The smeared link operator $U_i^{(n)}$ is constructed by maximizing 
\begin{equation}
{\rm Re\, Tr} [U_i^{(n+1)}(t, \vec{x}) V_i^{(n)\dagger}(t, \vec{x}) ]
, 
\label{eq:smearing}
\end{equation}
where
\begin{eqnarray}
V_i^{(n)}(t, \vec{x})
&\equiv &
\alpha U_i^{(n)}(t, \vec{x})
+\sum_{\pm j \neq i} U_j^{(n)}(t, \vec{x})
\nonumber\\
&& 
\times
U_i^{(n)}(t, \vec{x}+a\vec{e}_j) U_j^{(n)\dagger}(t, \vec{x}+a\vec{e}_i)
.
\end{eqnarray}
The optimal choices of $\alpha$ and $n$ for each $\beta$ are given in Table \ref{table:glueballmass}.
We compare in Fig. \ref{fig:effmass} the effective mass plots with the smeared and unsmeared glueball operators.
We see that the smeared operator requires much less imaginary time to form the plateau, and the statistical error is much smaller.
With the above setup, we found the glueball mass $m_\phi$ as shown in Table \ref{table:glueballmass} for $\beta = 2.1, 2.2, 2.3, 2.4$, and 2.5.
The largest uncertainty of $m_\phi$ is coming from the relation (\ref{eq:stringtensionNc}).

\begin{table}[htb]
\caption{
Smearing parameters $\alpha$ and $n$ used to optimize the 0$^{++}$ glueball operator.
The results of our calculations of 0$^{++}$ glueball masses (in lattice unit) are also shown.
}
\begin{ruledtabular}
\begin{tabular}{l|cccc}
$\beta$ & $\alpha$ & $n$ & $a m_\phi$ & $m_\phi [\Lambda]$ \\ 
\hline
2.1 & 16.0 & 3 & 1.853(13) & 5.21(39) \\
\hline
2.2 & 16.0 & 5 & 1.517(10) & 5.55(41) \\
\hline
2.3 & 10.0 & 7 & 1.241(6) & 5.75(54) \\
\hline
2.4 & 2.0 & 11 & 0.924(8) & 5.93(43) \\
\hline
2.5 & 2.0 & 27 & 0.696(6) & 6.32(46) \\
\end{tabular}
\end{ruledtabular}
\label{table:glueballmass}
\end{table}

\subsection{Nambu-Bethe-Salpeter amplitude\label{sec:NBS}}

\begin{figure*}[htbp]
\centering
\includegraphics[width=0.48\textwidth,clip]{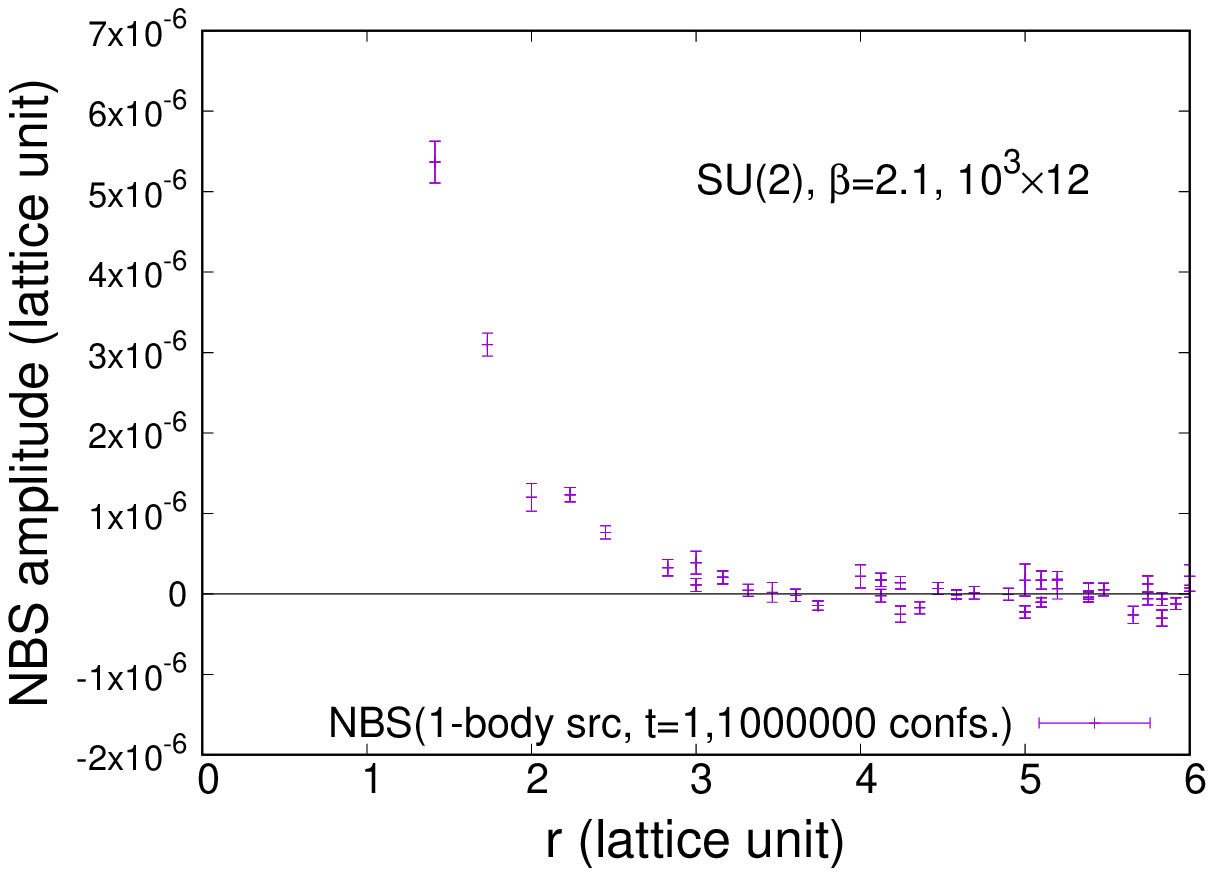}
\includegraphics[width=0.48\textwidth,clip]{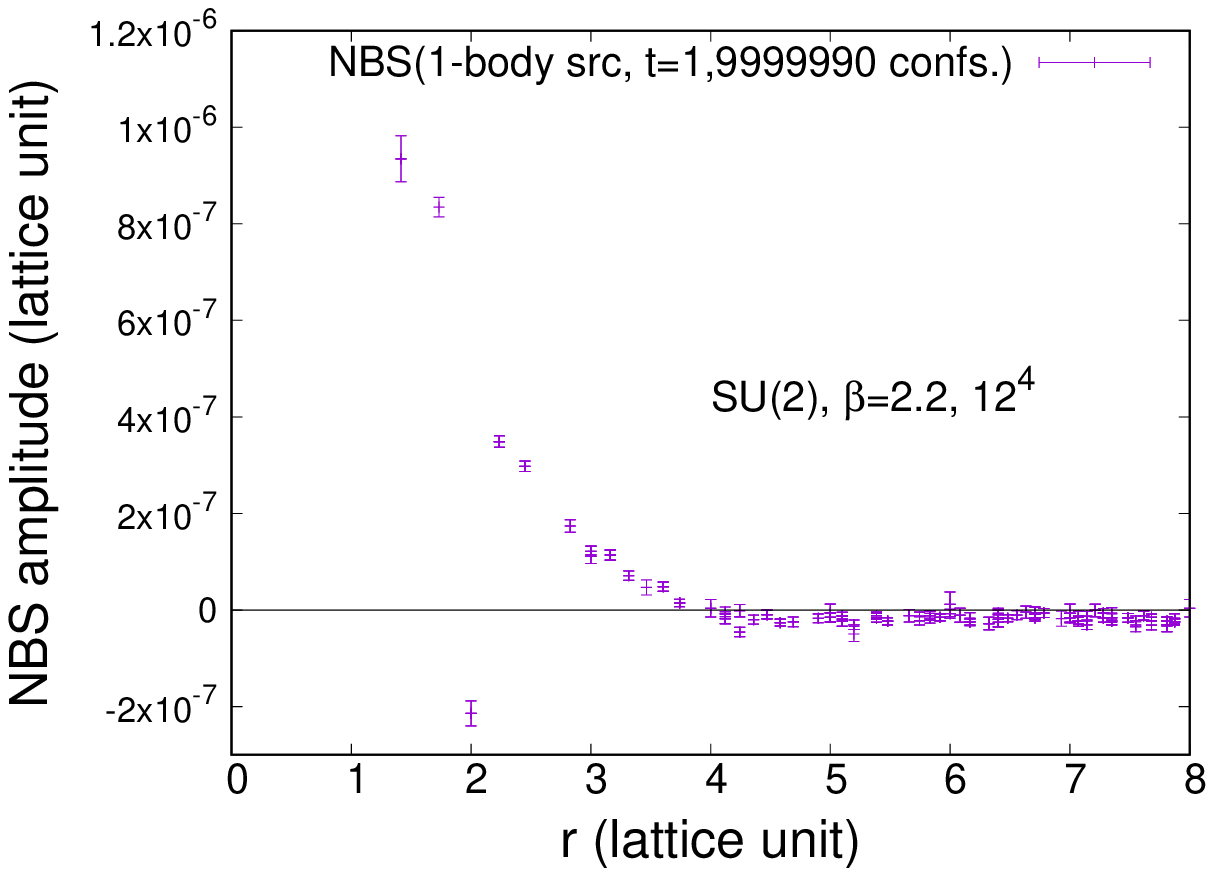}
\vspace{1mm}
\includegraphics[width=0.48\textwidth,clip]{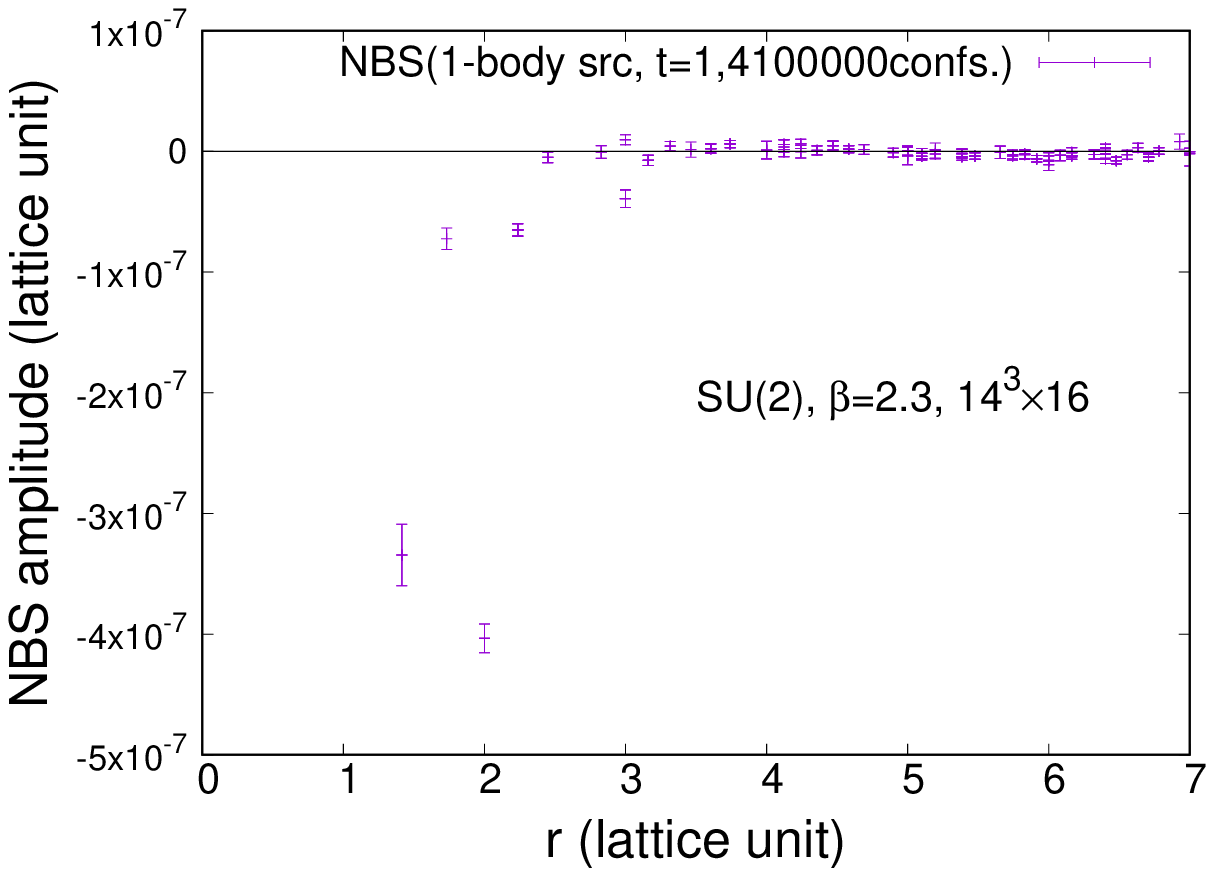}
\includegraphics[width=0.48\textwidth,clip]{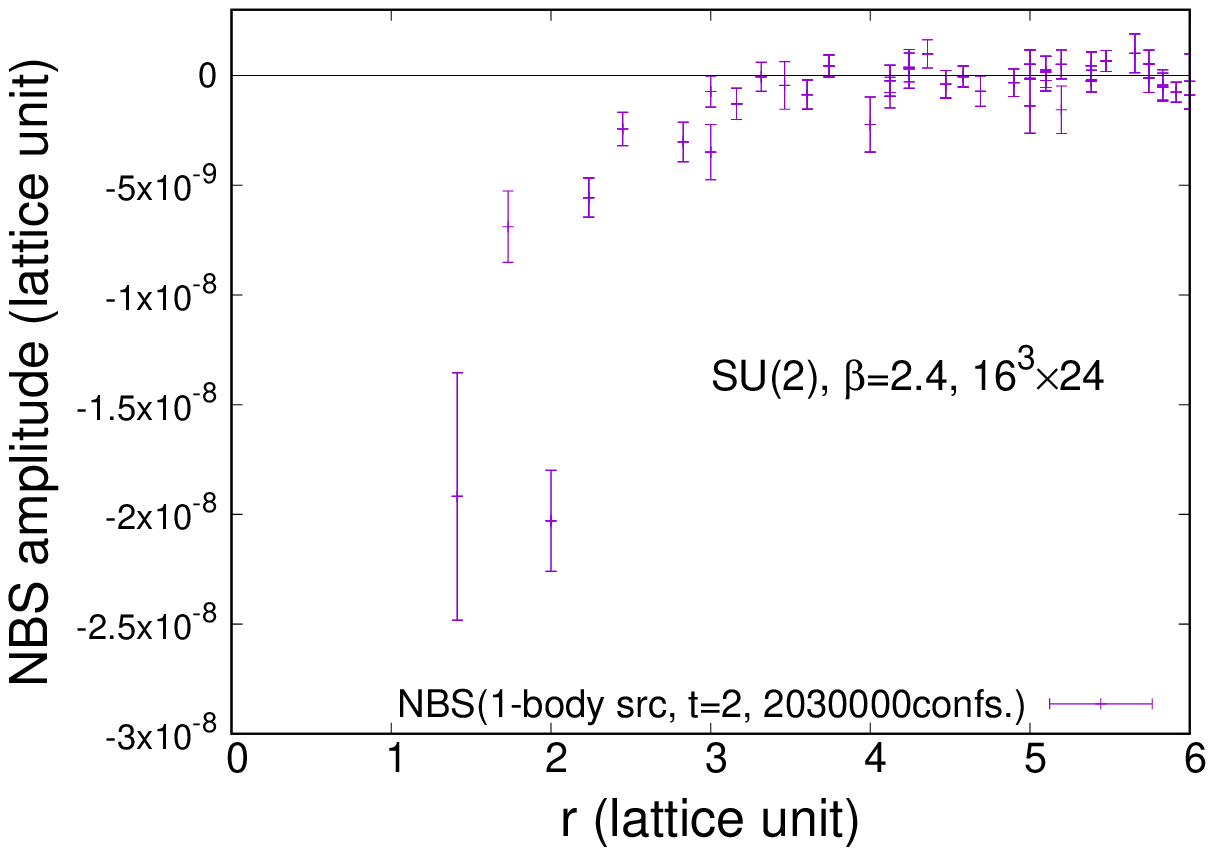}
\vspace{1mm}
\includegraphics[width=0.48\textwidth,clip]{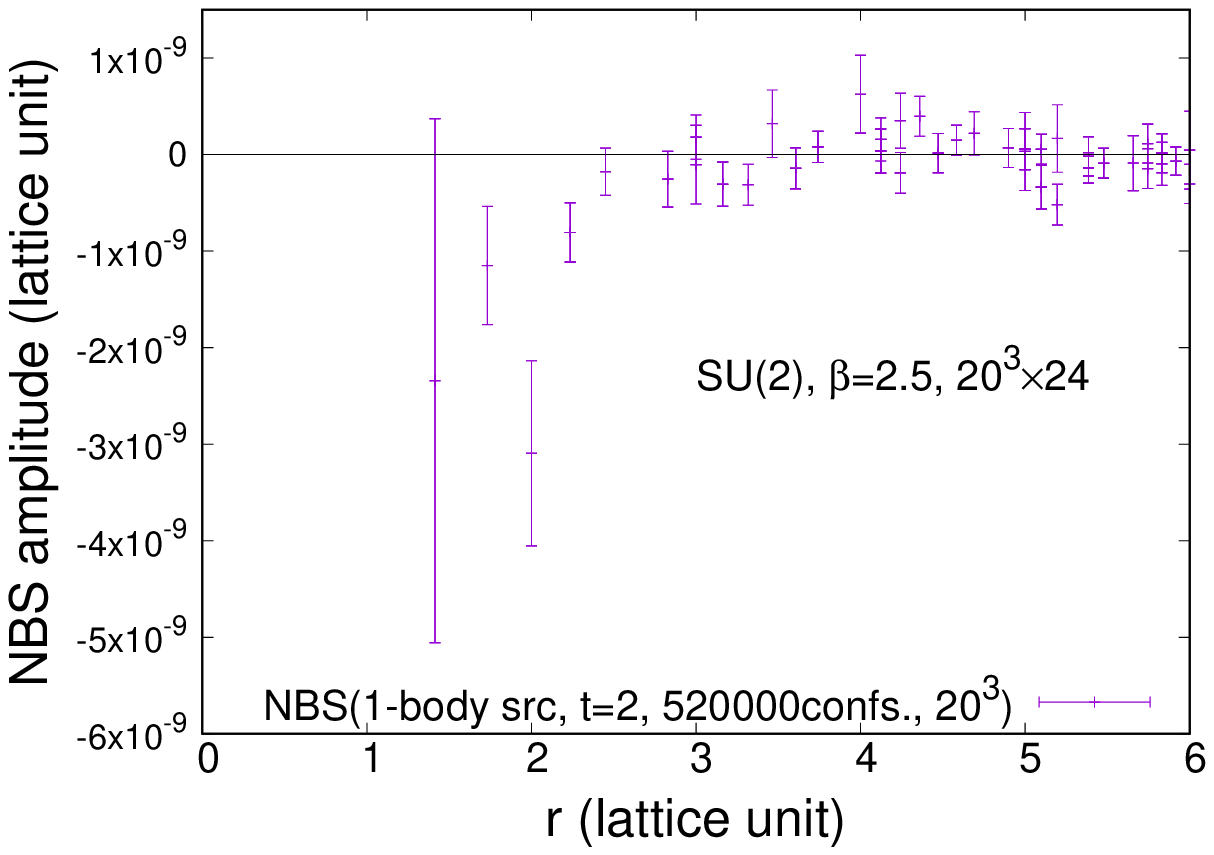}
\caption{\label{fig:su2_glueball_BS_1-body}
NBS amplitudes for the 1-body wall source calculations on lattice with $\beta =2.1, 2.2, 2.3, 2.4$, and 2.5.
}
\end{figure*}

\begin{figure}[htbp]
\centering
\includegraphics[width=0.48\textwidth,clip]{su2_beta2p4_glueball_BS_1-body.eps}
\vspace{1mm}
\includegraphics[width=0.48\textwidth,clip]{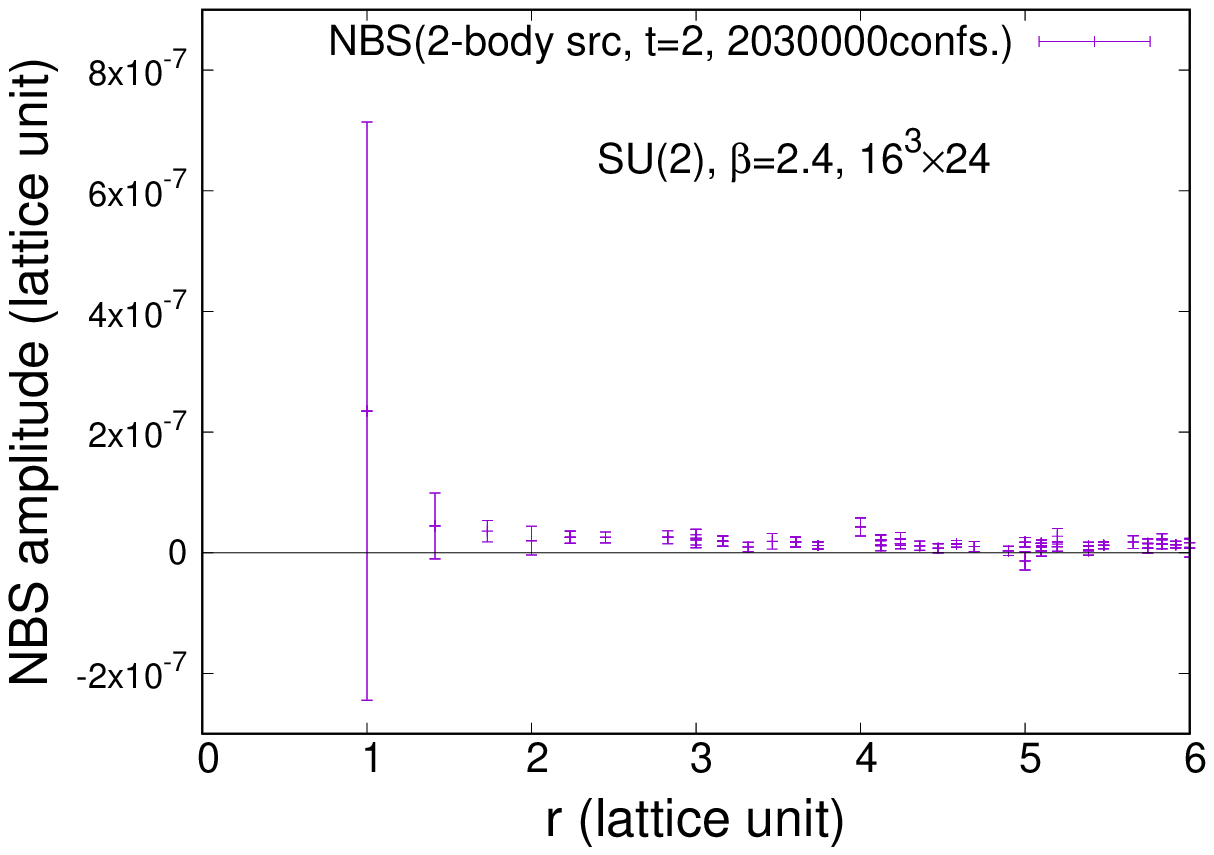}
\vspace{1mm}
\includegraphics[width=0.48\textwidth,clip]{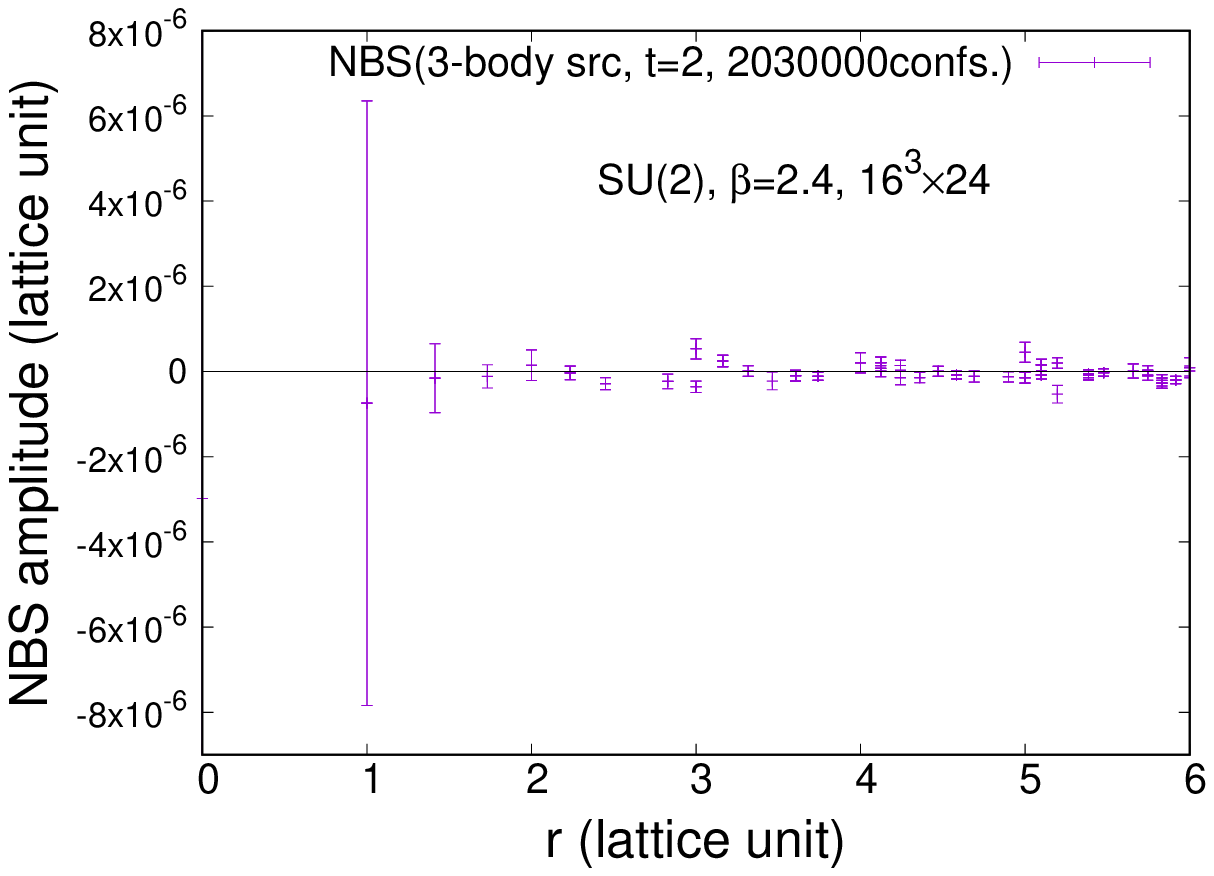}
\caption{\label{fig:su2_beta2p5_glueball_BS_n-body}
NBS amplitudes for the 1-, 2-, and 3-body wall source calculations on  $16^3 \times 24$ lattice with $\beta =2.4$.
}
\end{figure}

To extract the scattering cross section between two hadrons, we have to calculate the Nambu-Bethe-Salpeter (NBS) amplitude, defined as follows:
\begin{equation}
\Psi_{\phi \phi} 
(t,\vec{x}-\vec{y})
\equiv
\frac{1}{V} \sum_{\vec{r}} 
\langle 0 | T[\tilde \phi (t, \vec{x}+\vec{r})\tilde \phi (t, \vec{y}+\vec{r}) {\cal J}(0)] | 0 \rangle
.
\label{eq:NBS}
\end{equation}
Here ${\cal J}$ is the source operator which has the same quantum number as the two-glueball state, and it may be improved using the APE smearing.
However, the sink operators should not be smeared, since the nonlocality of the operator will affect the definition of the interglueball distance.
For the $0^{++}$ glueball, all $n$-body operator ($\tilde \phi^n$) ($n \in \mathbb{N}, n \neq 0$) may be chosen.
We also note that ${\cal J}$ also has expectation value, so we have to subtract it.
We can easily show that the removal at one side, either at the source or at the sink, is sufficient, since we have
\begin{eqnarray}
&& \hspace{-2em}
\langle 0 | T[ [{\cal O}_{\rm snk} (t, \vec{r}) -\langle {\cal O}_{\rm snk} (t, \vec{r})\rangle ][{\cal O}_{\rm src} (0) -\langle {\cal O}_{\rm src} (0)\rangle ]  ] | 0 \rangle
\nonumber\\
&=&
\langle 0 | T[ [{\cal O}_{\rm snk} (t, \vec{r}) -\langle {\cal O}_{\rm snk} (t, \vec{r})\rangle ] {\cal O}_{\rm src} (0) | 0 \rangle
\nonumber\\
&=&
\langle 0 | T[ {\cal O}_{\rm snk} (t, \vec{r}) [{\cal O}_{\rm src} (0) -\langle {\cal O}_{\rm src} (0)\rangle ]  ] | 0 \rangle
.
\label{eq:NBSdefinition}
\end{eqnarray}
For the computational convenience, we choose to remove the expectation value of the source ${\cal J}$.
We show in Fig. \ref{fig:su2_glueball_BS_1-body} the behavior of the NBS amplitude with 1-body source at $\beta = 2.1, 2.2, 2.3, 2.4$, and 2.5.
We see characteristic oscillations with a valley at each lattice unit.
This integer variation is suggesting us the relevance of nonzero angular momentum ($l\ge 4$) effects due to the lattice discretization.
At integer $r$, $l=4$ wave contributes, while other higher partial waves also contribute at other non-integer $r$.
This fact is the origin of the above integer oscillation.
We will see later how to remove this systematic effect when extracting the interglueball potential.

In Fig. \ref{fig:su2_beta2p5_glueball_BS_n-body}, we compare the same quantity between 1-, 2-, and 3-body sources for $\beta =2.4$.
We see that for the 1-body source calculation, the NBS amplitude becomes zero at large $r$, while it is finite for the cases of 2- and 3-body sources.
This is because the NBS amplitude with 1-body source cannot be splitted into two spatially separated correlators, while that with 2- and 3-body sources can.
We note that this fact does not occur if the expectation values of the source (and sink) operator are not correctly subtracted, and this proves that the cluster decomposition principle is correctly working with the definition (\ref{eq:NBSdefinition}).
Since the lattice calculation becomes noisier as the mass dimension of the operator increases, we will mainly discuss the NBS amplitude with the 1-body source.
The correlator (\ref{eq:NBS}) is purely gluonic, and the statistical error is significant in the lattice calculation.
To improve the accuracy, we use all space-time symmetries (space-time translation and cubic rotation) to effectively increase the statistics.

\subsection{\label{sec:volume}Finite volume effect}

We now inspect the finite volume effect.
In looking at the damping of each NBS amplitude of Fig. \ref{fig:su2_glueball_BS_1-body}, we see that the signal is correctly becoming zero consistent before reaching the half of the lattice spatial length.

\begin{figure}[htb]
\includegraphics[width=8.5cm]{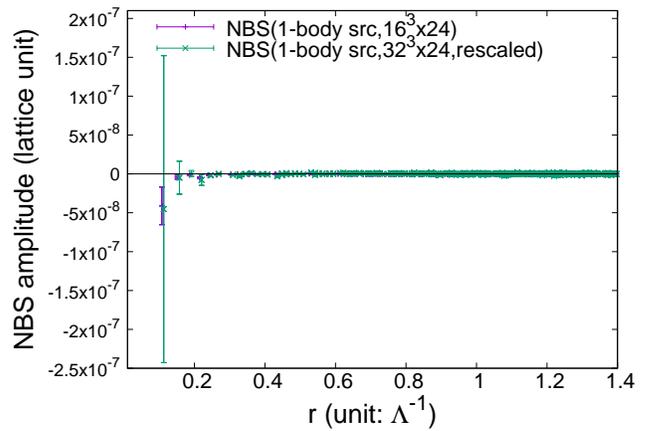}
\caption{\label{fig:BSfinitevolume}
Comparison of the calculations of the NBS amplitude using the 1-body wall source with the volumes $16^3\times 24$ and $32^3\times 24$ ($\beta =2.5$).
The NBS amplitude was calculated with 1045000 and 126000 configurations for the $16^3\times 24$ and $32^3\times 24$ lattices, respectively.
}
\end{figure}

Another important point to be discussed in the context of the finite volume effect is the vacuum fluctuation of gluonic operators.
Let us compare the lattice calculations with two different volumes $16^3\times 24$ and $32^3\times 24$ for $\beta=2.5$.
We plot in Fig.~\ref{fig:BSfinitevolume} the results of the NBS amplitude (1-body wall source) calculated with approximately the same statistics, which are considered to be proportional to the volume thanks to the use of translational symmetries.
We see that the results are in agreement, although the statistical error is large for the case of $32^3\times 24$ lattice.
It is actually known that the noise enhances when distant gluonic correlations (disconnected insertions) contribute to the correlator, since such contribution increases with the volume, as we chose the wall source.
In the next section, we precisely use this fact to reduce the statistical uncertainty, by removing the uncorrelated contribution.

\subsection{The cluster-decomposition error reduction technique}

\begin{figure}[htb]
\includegraphics[width=8.5cm]{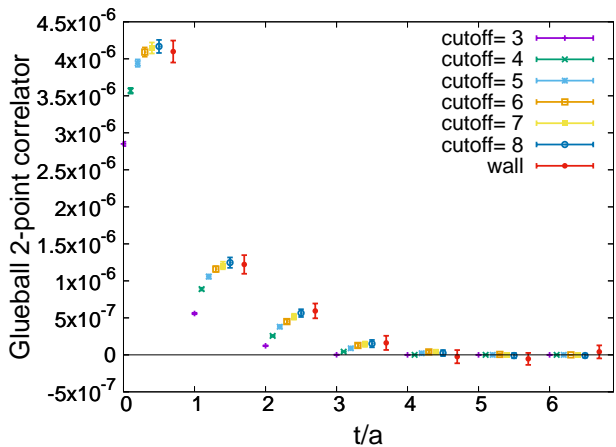}
\caption{\label{fig:2ptCD}
CDERT applied to the glueball two-point function calculated with 100 configurations and the smeared operator on $16^3 \times 24$ lattice with $\beta =2.4$.
We see that the correlator saturates at the cutoff $\rho =7$ (lattice unit), and further increase of the cutoff enlarges the statistical error bar.
}
\end{figure}

As we saw in Section \ref{sec:volume}, the contribution to the correlator from gluonic operators that are four-dimensionally well separated yields statistical fluctuation which accumulates with the increase of the volume.
This is because these operators have expectation values and they are fluctuating even when they are isolated.
The idea then came to remove these meaningless fluctuations originating from the distant positions of the operators, and just keep the contribution from the true correlation with closely located interpolating fields.
Since the YMT has a mass gap, the correlation is exponentially suppressed in the Euclidean space, so that we can set a four-dimensional cutoff which removes the contribution from the uncorrelated region \cite{Liu:2017man}.
This is also an important technical application of the cluster decomposition principle.
We plot in Fig. \ref{fig:2ptCD} the example of the calculation of the glueball two-point correlator using the above mentioned method, the cluster-decomposition error reduction technique (CDERT).
We found that, for $\beta = 2.4$, the systematic error is less than the statistical one with the cut $\rho = 7$ (lattice unit).

\begin{figure}[htb]
\includegraphics[width=8.5cm]{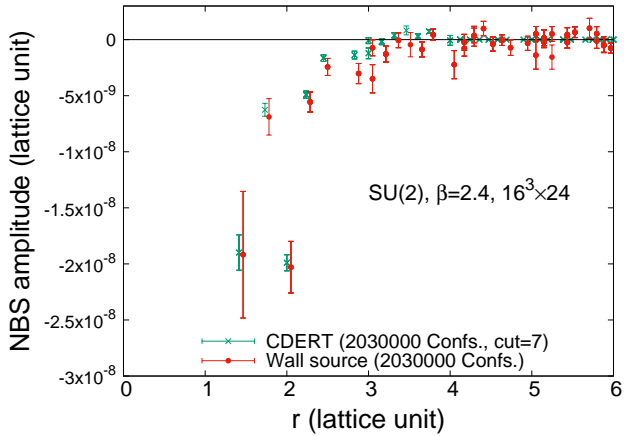}
\caption{\label{fig:BS}
Glueball NBS amplitude with 1-body source (${\cal J} = \tilde \phi$) obtained by applying the CDERT with the cutoff $\rho = 7$ (lattice unit) on $16^3 \times 24$ lattice at $\beta=2.4$.
We compare it with the wall source calculation to visualize the improvement of the signal.
}
\end{figure}

In this work, we apply the CDERT used in Ref. \cite{Liu:2017man} to the calculation of the NBS amplitude on $16^3 \times 24$ lattice with $\beta =2.4$, for which the computational cost was optimal (for $\beta=2.1, 2.2, 2.3$, increasing the number of configurations was more efficient to reduce the statistical error, while the application of the CDERT was too costly for $20^3 \times 24$ lattice with $\beta =2.5$).
We set cutoffs to the relative four-dimensional distances between the source operator and the sink ones.
The NBS amplitude with the 1-body source is then calculated as
\begin{eqnarray}
\Psi'_{\phi \phi} (t,\vec{x}-\vec{y})
&=&
\frac{1}{V} \sum_{\vec{r}} 
\sum_{\vec r_{\rm src}\in C (t, \vec{x}+\vec{r} ) \bigcup C (t,\vec{y}+\vec{r} )} 
\nonumber\\
&& 
\langle 0 | T[\tilde \phi (t, \vec{x}+\vec{r})\tilde \phi (t, \vec{y}+\vec{r}) \tilde \phi (0, \vec r_{\rm src}  )] | 0 \rangle
,
\nonumber\\
\end{eqnarray}
where $C (t, \vec{v} )$ is the 3-dimensional projection of the 4-dimensional hypersphere with the center $(t, \vec{v} )$ and with the radius $\rho$ onto the $t=0$ (3-dimensional) hyperplane.
The cutoffs are simultaneously applied to both glueball operators of the sink, which means that the wall source is changed to a source with a restricted region where two spheres of the cutoffs overlap. 
We note that this simultaneous application of the cutoff to the above two pairs of operators upsets the possibility to reduce the computational cost using the Fourier transform, which was very efficient in the case of two-point functions \cite{Liu:2017man}.
In the case of $SU(2)$ YMT, the calculation of the NBS amplitude with the CDERT is the most computationally costly step of this work.
We also apply the same cutoff to the distance between the two interpolating fields of the sink, but this manipulation is just equivalent to not considering the radial plot of the NBS amplitude for the radius beyond the cutoff.
In Fig. \ref{fig:BS}, we plot the result of the application of the CDERT to the NBS amplitude calculated on $16^3 \times 24$ lattice at $\beta = 2.4$ with the cutoff $\rho = 7$ (lattice unit).
We see that the CDERT could successfully reduce the statistical error by more than a factor of two, keeping consistency with the wall source calculation.

\subsection{Problems with the direct method\label{sec:directmethod}}

\begin{figure}[htb]
\includegraphics[width=8.5cm]{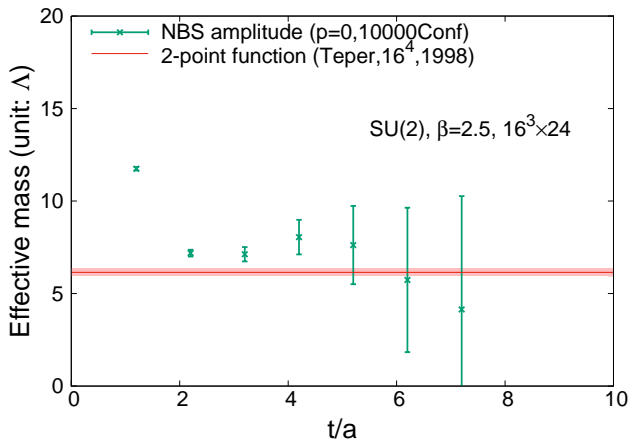}
\caption{\label{fig:su2_beta2p5_glueball_BSFT_zero_mom}
Effective mass plot of the glueball NBS amplitude in the momentum space, with 1-body source (${\cal J} = \tilde \phi$) on $16^3 \times 24$ lattice with $\beta = 2.5$.
We see that the energy of the system saturates at the single glueball mass.
}
\end{figure}

We now extract the scattering phase shift.
It may be calculated by Fourier transforming the NBS amplitude and by inspecting the momentum modulation of the energy of the two-glueball system (the so-called L\"{u}cher's method) \cite{Luscher:1990ux}.
This approach is expected to be applicable when there are no states with smaller energy or when the quantum number of the two-body system forbids transition to lighter states, and it has been applied to many hadronic systems  \cite{Beane:2005rj,Beane:2006gj,Beane:2007xs,Beane:2010em,Beane:2010hg,Beane:2012vq,Briceno:2016mjc,Francis:2018qch}.
However, in the case of the two-glueball state, it mixes with the single glueball state.
Indeed, the NBS amplitude forms a plateau with the effective mass of a single glueball (see Fig. \ref{fig:su2_beta2p5_glueball_BSFT_zero_mom}).
It is, of course, possible to remove the contribution from the one-glueball state by hand or by diagonalizing the source operator, but we have to keep in mind that the glueball spectrum has other states and resonances with energy close to the two-body threshold.
The extraction of the two-glueball scattering in the direct method is therefore very challenging.

\subsection{HAL QCD method}

An alternative approach to calculate the scattering phase shift is to indirectly calculate it via the interglueball potential obtained from the HAL QCD method \cite{Ishii:2006ec,Aoki:2009ji}.
The object is to extract the nonlocal potential $U({\vec r},{\vec r}')$ by using the fact that the NBS amplitude $\Psi_{\phi \phi} (t,{\vec r})$ obeys the following time-independent Schr\"{o}dinger-like equation:
\begin{equation}
\frac{1}{m_\phi} \nabla^2
\Psi_{\phi \phi} (t,{\vec r}) 
=
\int d^3r' U({\vec r},{\vec r}')
\Psi_{\phi \phi} (t,{\vec r}') 
.
\label{eq:standardHAL}
\end{equation}
By taking a sufficiently large $t$ (we call this ``taking the ground saturation''), the potential can be extracted.
It is easily possible to realize large $t$ with large time discretization (lattice spacing), i.e. small $\beta$.
In our calculation, we use the time-independent formalism to extract the interglueball potential for $\beta = 2.1$ and 2.2 [Eq. (\ref{eq:standardHAL}) will not be used exactly in its form, since we have to subtract the centrifugal force, as seen below].

The above method is unfortunately difficult to apply to cases where the contamination from excited states is important.
To overcome this problem, the HAL QCD method was then extended to the following time-dependent formalism \cite{HALQCD:2012aa}
\begin{equation}
\Biggl[
\frac{1}{4m_\phi} \frac{\partial^2}{\partial t^2}-\frac{\partial}{\partial t} + \frac{1}{m_\phi} \nabla^2
\Biggr]
R(t,{\vec r})
=
\int d^3r' U({\vec r},{\vec r}')R(t,{\vec r}')
,
\end{equation}
where $R(t,{\vec r}) \equiv \frac{\Psi_{\phi \phi} (t,{\vec r}) }{e^{-2m_\phi t}}$.
Here we have to choose $t$ so that $1/t$ is less than the inelastic threshold $E_T=3m_\phi - 2m_\phi= m_\phi$, if one wants to study the elastic scattering.
The elasticity is an important feature of the HAL QCD method, since it makes the physics essentially nonrelativistic.
The potential should then be local and central, $U({\vec r},{\vec r}') \approx V_{\phi \phi} (\vec{r}) \delta (\vec{r}-\vec{r}')$, to a good approximation.
We then have
\begin{equation}
V_{\phi \phi} (\vec{r})
=
\frac{1}{R(t,{\vec r})}
\Biggl[
\frac{1}{4m_\phi} \frac{\partial^2}{\partial t^2}-\frac{\partial}{\partial t} + \frac{1}{m_\phi} \nabla^2
\Biggr]
R(t,{\vec r})
.
\label{eq:HALpotential}
\end{equation}
The above extension to the time-dependent formulation has the crucial advantage that we do not need the ground state saturation for extracting the potential \cite{HALQCD:2012aa}, and it is now well established after intense discussions \cite{Iritani:2016jie,Iritani:2017rlk,Yamazaki:2017gjl,Aoki:2017yru,Iritani:2018vfn}.
In particular, the glueball correlators are in general very noisy, so the use of this method is almost mandatory for large $\beta$ if one wants to keep good statistical accuracy.
In addition, the potential handled in the HAL QCD method does not depend on the renormalization scale \cite{Ishii:2006ec,Aoki:2009ji}.
We, however, have to keep in mind that the potential is not an observable, and it may depend on the choice of the operators.

As seen in Sec. \ref{sec:NBS}, the NBS amplitude calculated on lattice might be contaminated by higher partial waves with angular momentum $l \ge 4$.
The HAL QCD Collaboration resolved this problem by using Misner's method \cite{Miyamoto:2019jjc}.
This consists of taking the weighted average of the NBS amplitude over a thin interval of distance between hadrons.
However, for the case of the glueballs, the signal of the potential is spatially damping very fast and only data points close to the origin may be used, so we will not have enough points in a thin interval to take the weighted average.
In our study, we resolve this problem instead by simply subtracting the centrifugal force which is giving the leading contribution from finite angular momenta in the Schr\"{o}dinger equation.
The relation between the potential and the NBS amplitude is then
\begin{equation}
V_{\phi \phi} (\vec{r})
=
\frac{1}{R(t,{\vec r})}
\Biggl[
\frac{1}{4m_\phi} \frac{\partial^2}{\partial t^2}-\frac{\partial}{\partial t} + \frac{\nabla^2}{m_\phi}
+\frac{(\vec{r} \times \vec{\nabla})^2}{2m_\phi r^2}
\Biggr]
R(t,{\vec r})
.
\label{eq:improvedHALpotential}
\end{equation}
In the case of the glueball, the signal is also very quickly damping in the temporal direction, so the removal of the centrifugal force, which should have a large effect at short distance, is also expected to improve the quality of the potential at small imaginary time, without waiting for the damp of the centrifugal one.
Moreover, in our study, we are interested in the low energy scattering between DM, so it is fortunate that we only have to consider the s-wave scattering.
We also note that the subtraction of the centrifugal force from the kinetic term yields
\begin{equation}
\frac{\nabla^2}{m_\phi}
+\frac{(\vec{r} \times \vec{\nabla})^2}{2m_\phi r^2}
=
\frac{\vec{r} \cdot \vec{\nabla}}{m_\phi r^2}
-\sum_{a,b, =1}^3\frac{r_a r_b \nabla_a \nabla_b}{2 m_\phi r^2}
,
\end{equation}
which is not containing the Laplacian $\nabla^2$ anymore.
This means that the second derivative disappears (we instead have a product of first derivatives $\nabla_a \nabla_b$), and consequently the nonlocality is reduced.
This will permit us to extract the potential with data points close to the origin, without being annoyed by the contact term of glueball operators.

\begin{figure}[htb]
\includegraphics[width=8.5cm]{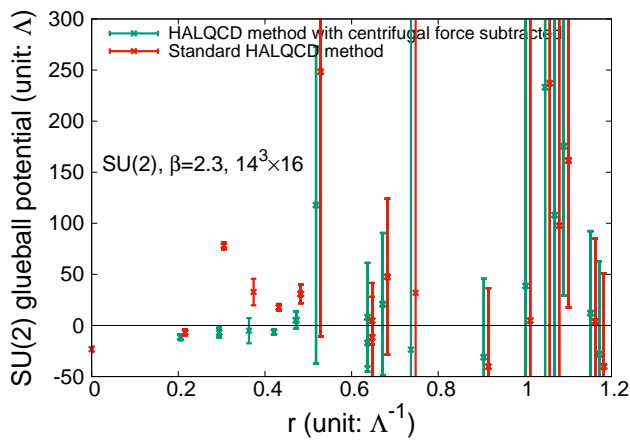}
\caption{\label{fig:su2_beta2p3_glueball_potential_centrifugal}
Comparison of the interglueball potentials extracted using the HAL QCD method with and without the subtraction of centrifugal force ($14^3 \times 16$ lattice, $\beta =2.3$).
The potential becomes attractive after the subtraction in the region around $r \sim 0.3$.
}
\end{figure}

In Fig. \ref{fig:su2_beta2p3_glueball_potential_centrifugal}, we compare the interglueball potentials extracted using the HAL QCD method with and without the subtraction of centrifugal force.
We see that the potential is repulsive before the subtraction of the centrifugal force, and turns attractive after the improvement of Eq. (\ref{eq:improvedHALpotential}), which shows that this procedure is very efficient in removing the systematics due to $l\ge 4$ effects.
We also remark that the error bars are very large in the region $r \ge 0.5 \, \Lambda^{-1}$.
This is due to the fact that the potential is calculated by dividing by the NBS amplitude [see Eqs. (\ref{eq:HALpotential}) and (\ref{eq:improvedHALpotential})] which is zero in the long range for the case of the 1-body source, due to the cluster decomposition principle (see Figs. \ref{fig:su2_glueball_BS_1-body}, \ref{fig:su2_beta2p5_glueball_BS_n-body}, and \ref{fig:BS}).

\section{\label{sec:results}Results and analysis}

\subsection{\label{sec:interglueballpotential}Interglueball potential}

We calculate the interglueball potential for $\beta = 2.1, 2.2, 2.3, 2.4$, and 2.5 and superpose the results without weight.
This manipulation is allowed since the potential obtained in the HAL QCD method is independent of the choice of the renormalization scale.
For $\beta = 2.3, 2.4$, and 2.5, the interglueball force is extracted using the time-dependent formalism improved by removing the centrifugal force (\ref{eq:improvedHALpotential}).
For $\beta = 2.1$ and 2.2, the time interval is large, and the use of Eq. (\ref{eq:improvedHALpotential}) introduces sizable systematics due to the discretization through the time derivative.
In this work, we therefore employ the standard HAL QCD method (\ref{eq:standardHAL}) improved by subtracting the centrifugal force to extract the potential from the lattice data at $\beta = 2.1$ and 2.2.

\begin{figure}[htb]
\includegraphics[width=8.5cm]{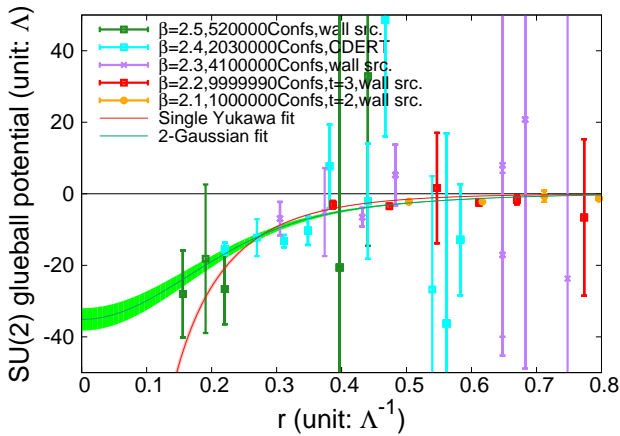}
\caption{\label{fig:potential}
Interglueball potential calculated on lattice in $SU(2)$ YMT at $\beta = 2.1, 2.2, 2.3, 2.4$, and 2.5.
The fits with the Yukawa (\ref{eq:Yukawafit}) and 2-Gaussian (\ref{eq:2-Gaussianfit}) fitting forms are also displayed.
The colored band denotes the uncertainty (statistical and systematic).
For each $\beta$, we do not displayed data points at $r =0, 1$, and $r \ge 4$ (lattice unit) because we did not use them in the fit.
}
\end{figure}

We plot in Fig. \ref{fig:potential} the result of our calculation.
We see from our result that the interglueball potential is attractive.
Here we note that we removed the data points at $r=0$ and $r=1$ (in lattice unit).
This is because the glueball operator we used is a superposition of plaquettes so as to form a 3-dimensional cube of spread with one lattice unit [see Eq. (\ref{eq:glueballop})], and the product of two glueball operators at the same or consecutive spatial points may generate an operator made of overlapping plaquettes, which is a single glueball operator.
This problem was also encountered in the direct method (see Sec. \ref{sec:directmethod}).

Let us now fit the potential calculated on lattice.
Here we choose two fitting forms, i.e. the Yukawa function 
\begin{equation}
V_Y(r) = V_Y^{(1)} \frac{e^{-m_\phi r}}{4 \pi r},
\label{eq:Yukawafit}
\end{equation}
and the two-range Gaussian (2-Gaussian) function 
\begin{equation}
V_{G}(r) = V_{G}^{(1)} e^{-\frac{(m_\phi r)^2}{8}} +V_G^{(2)} e^{-\frac{(m_\phi r)^2}{2}}.
\label{eq:2-Gaussianfit}
\end{equation}
After the fit, we find 
\begin{eqnarray}
V_Y(r) 
&= &
-231(8) \frac{e^{-m_\phi r}}{4 \pi r} \ \ \ (\chi^2 / {\rm d.o.f.} = 1.3)
,
\label{eq:fittedYukawa}
\\
V_{G}(r) 
&=& 
-8.5(0.5) \Lambda e^{-\frac{(m_\phi r)^2}{8}} -26.6(2.6) \Lambda e^{-\frac{(m_\phi r)^2}{2}}
\nonumber\\
&& \hspace{8em}
(\chi^2/ {\rm d.o.f.} = 0.9)
.
\label{eq:fitted2-Gaussian}
\end{eqnarray}
The results are plotted in Fig. \ref{fig:potential}.
Both fits have $\chi^2$ close to one, which shows that they work reasonably well.

\subsection{\label{sec:glueballlagrangian}0$^{++}$ glueball effective Lagrangian}

\begin{figure}[htb]
\includegraphics[width=3cm]{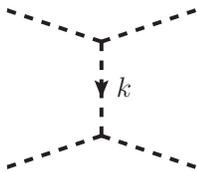}
\caption{\label{fig:Yukawa_potential}
One-glueball exchange process in the t-channel generated by the glueball trilinear interaction, with the exchanged momentum $k$.
}
\end{figure}

Let us try to analyze our result in terms of a simple scalar effective field theory.
The general renormalizable effective theory of the $0^{++}$ glueball is given by the following trilinear + $\phi^4$ Lagrangian  
\begin{equation}
{\cal L}_\phi
=
\frac{1}{2} ( \partial^\mu \phi )^2
-\frac{m_\phi^2}{2} \phi^2
-\frac{A}{3!} \phi^3
-\frac{\lambda}{4!} \phi^4
.
\end{equation}
The above trilinear interaction (term with $A$) generates an attractive Yukawa potential (see Fig. \ref{fig:Yukawa_potential}) which has the longest range, and explains well our lattice result.
By matching the one-glueball exchange process with the nonrelativistic Yukawa potential, we have $|A| = 2 m_\phi \sqrt{-V_Y^{(1)}} \approx 190 \Lambda$.
The $\phi^4$ interaction may be repulsive depending on the sign of the coupling $\lambda$, but this one is giving a delta function potential in the continuum theory, and it is difficult to fit it directly from lattice data.

The trilinear coupling $A$ of the lightest 0$^{++}$ glueball has actually been extracted in $SU(3)$ lattice YMT \cite{deForcrand:1984eeq} using L\"{u}scher's finite volume method \cite{Luscher:1990ux,Luscher:1983rk}, and it was obtained $\frac{3A^2}{ 16 \pi m_\phi^2} = 155\pm 45$, which is comparable to our result of $SU(2)$ YMT $\frac{3A^2}{ 16 \pi m_\phi^2} \approx 60$.
This is also consistent with the calculation of the strong coupling expansion $\frac{3A^2}{ 16 \pi m_\phi^2} \approx 120$ \cite{Munster:1984zf}.
As an aside, we note that the glueball mass $m_\phi^2$, the trilinear coupling $A$, and the $\phi^4$ coupling $\lambda$ scale as $O(N_c^0)$, $O(N_c^{-1})$, and $O(N_c^{-2})$, in the large $N_c$ limit, respectively. 
This results in the scaling of the interglueball cross section $\sigma_{\phi \phi} = O(N_c^{-4})$.

It is also possible to derive the glueball effective Lagrangian by using the conformal invariance and by supposing that the glueball field operator is the trace anomaly, which explicitly breaks the scale (conformal) symmetry at the quantum level.
In the low energy limit, it has the following unique form \cite{Ellis:1970yd,deAlfaro:1976vlx,Collins:1976yq,Schechter:1980ak,Salomone:1980sp,Migdal:1982jp,Ellis:1984jv,Gomm:1984zq,Gomm:1985ut,Birse:1994ar,Kharzeev:2008br}
\begin{equation}
{\cal L}_{\phi}
=
\frac{1}{2} e^{2\frac{\phi}{f_\phi}} (\partial^\mu \phi ) (\partial_\mu \phi )
-H_0 
\left(
\frac{1}{4} - \frac{\phi}{f_\phi}
\right)
e^{4\frac{\phi}{f_\phi}}
,
\label{eq:glueballEFT}
\end{equation}
where the decay constant $f_\phi$ and the vacuum expectation value $H_0$ are the only two free parameters.
With this effective theory, the glueball mass is given as $m_\phi = \sqrt{-4H_0 /f_\phi^2}$.
To completely determine the low energy effective Lagrangian of the glueball, another constraint is required.
Here we match the one-glueball exchange process generated by Eq. (\ref{eq:glueballEFT}) with the nonrelativistic Yukawa potential (\ref{eq:fittedYukawa}) (see Appendix \ref{sec:glueballEFT} for derivation).
After matching we have
\begin{eqnarray} 
|f_\phi |
& \approx&
1.4 \Lambda
,
\label{eq:decayconstant}
\\
H_0
&\approx &
-18 \Lambda^4
.
\label{eq:energydensity}
\end{eqnarray} 
From the above discussion, we could obtain the dependence of the low energy glueball effective Lagrangian (\ref{eq:glueballEFT}) on the scale parameter $\Lambda$ of $SU(2)$ YMT.
This is actually the first ab initio derivation of the low energy dynamics of $SU(2)$ YMT, and we expect it to be applied in the predictions of important low energy observables.
Since the deconfinement transition of $SU(2)$ YMT occurs at the critical temperature $T_c \approx 1\times \Lambda$ \cite{Lucini:2003zr,Lucini:2012gg}, which is much smaller than the glueball mass $m_\phi \sim 6 \Lambda$, it is even possible to predict observables near $T_c$ using the above low energy glueball Lagrangian without large corrections.
Potentially interesting quantities to be evaluated are the DM relic density or the gravitational wave background which are generated at the phase transition.
We also note that $f_\phi \propto N_c $ and $H_0 \propto N_c^2$, so qualitative extrapolation of Eqs. (\ref{eq:decayconstant}) and (\ref{eq:energydensity}) to $N_c \ge 3$ is possible.

\subsection{Dark matter cross section and constraint on $\Lambda$}

We now derive the interglueball cross section by first solving the s-wave Schr\"{o}dinger equation with the fitted potentials (\ref{eq:fittedYukawa}) and (\ref{eq:fitted2-Gaussian})
\begin{eqnarray}
\Biggl(
\frac{\partial^2}{\partial r^2} 
+k^2
-m_\phi V(r)
\Biggr)
\phi(r)
=0
,
\label{eq:scatteringequation}
\end{eqnarray}
and extract the scattering phase shift $\delta(k)$ at vanishing momentum $k \to 0$, which is precisely the relevant kinematics for the DM scattering.
Asymptotically, the solution of Eq. (\ref{eq:scatteringequation}) behaves as
$\phi(r) \propto \sin [kr+\delta(k)]$.
The quantity we have to calculate is then
\begin{equation}
\sigma_{\phi \phi} 
=
\lim_{k\to 0}
\frac{4 \pi}{k^2} \sin^2 [\delta(k)]
.
\end{equation}
We finally obtain the following scattering cross sections for the two fitting forms we used:
\begin{eqnarray}
\sigma_{\phi \phi} 
&=& 
(2.5 - 4.7) \Lambda^{-2}  \ \ \  ({\rm Yukawa})
, 
\\
\sigma_{\phi \phi} 
&=& 
(14 - 51) \Lambda^{-2} \ \ \  (\mbox{2-Gaussian})
,
\end{eqnarray}
where the ranges correspond to the statistical error bar.
The systematic error is just estimated by taking the difference between the two results.
Finally, the interglueball scattering cross section in the $SU(2)$ YMT is
\begin{eqnarray}
\sigma_{\phi \phi} 
= 
(2 - 51) \Lambda^{-2} \ \ \ {\rm (stat.+sys.)}
.
\label{eq:gblimit}
\end{eqnarray}
Here we could see some enhancement of $\sigma_{\phi \phi} $ for the case of 2-Gaussian fit, which might be due to the existence of a resonance near the two-glueball threshold.
Its determination will be required for the reduction of the systematic error in the future.

We can then derive the constraint on $\Lambda$ from observational data.
By equating the upper bound on the DM cross section obtained from the galactic collision \cite{Harvey:2015hha}
\begin{equation}
\sigma_{\phi \phi} / m_{\phi} < 0.47 \, {\rm cm }^2 /{\rm g}
,
\end{equation}
we finally have
\begin{eqnarray}
\Lambda
>
60
\,{\rm MeV}
.
\end{eqnarray}
In this work, we do not discuss the lower bound on $\Lambda$ which might be set by inspecting the astrophysics at the scale below kpc, which has yet no consensus \cite{Spergel:1999mh,BoylanKolchin:2011dk,Brooks:2012ah,Chan:2015tna,Tollet:2015gqa,Hayashi:2015maa,Hayashi:2015yfa,Kaplinghat:2015aga,Behroozi:2016mne,Wetzel:2016wro,Elbert:2016dbb,Fitts:2016usl,Kim:2017iwr,Bullock:2017xww,Massey:2017cwf,vanDokkum:2018vup,1805.00484,Behroozi:2019kql,Allen:2019rzj,Harvey:2018uwf,1901.05973,Errani:2019sey}.

Using the large $N_c$ argument, we can qualitatively extend the discussion to all $N_c$'s.
As we saw in Section \ref{sec:glueballlagrangian}, the cross section scales as $1/N_c^4$, while the mass of the $0^{++} $glueball is constant at large $N_c$.
The lower limit of the scale parameter is then extended to $N_c \ge 3$ as
\begin{eqnarray}
\Lambda_{N_c}
>
60
\left( \frac{2}{N_c} \right)^{\frac{4}{3}}
\,{\rm MeV}
.
\end{eqnarray}
We note that the contribution of nonplanar diagrams to $\sigma_{\phi \phi}$ is of $O(N_c^{-6})$ so higher order corrections in $1/N_c$ expansion are not small.
To control the systematics down to the percent level, we have to calculate the interglueball cross section up to $N_c = 10$.

\section{Conclusion}

In this paper, we calculated the interglueball scattering cross section using the HAL QCD method, and derived the constraint on the scale parameter of $SU(2)$ YMT from the observational data of galactic collision.
Theoretically, the glueball DM is a natural conception because the theory does not depend on any massive parameters except the scale parameter.
It is also part of the nonabelian gauge theory with heavy vectorlike fermions.
The other dark gauge theories have more or less hierarchical problems, thus proving the attractiveness of the YMT.

In our work, the use of the HAL QCD method was almost mandatory, since this allowed us to take data with small imaginary time, which was crucial for extracting physical information from the very noisy glueball correlators.
We also used the CDERT at $\beta = 2.4$ which was shown to be efficient in the improvement of the signal.
Another important feature is the subtraction of the centrifugal force which removed the systematic effect due to nonzero angular momenta, which was crucial to correct the repulsive potential to an attractive one.
This attraction is consistent with the one-glueball exchange picture which is the leading contribution in the long range region.

We could also determine the low energy glueball effective theory by matching the one-glueball exchange process with the fitted Yukawa potential.
This effective Lagrangian is the unique form according to the Ward identity, and contains only two low energy constants.
Since the glueball mass is much larger than the temperature of deconfinement transition, our glueball effective theory is expected to be applicable even just below this temperature, and predictions of other observables such as the DM relic density are possible.
Determining the glueball effective field theory and its low energy constants may also give us an important insight into other fields such as the conformal field theory or hadron physics.
If again the large $N_c$ expansion holds, the determination of the cross section at a sufficiently large $N_c$ would probably mean the quantification of the conformal physics.

In our work, we assumed the local potential, but the systematics due to the nonlocality has to be checked in the future, since we defined the glueball operator with plaquettes which have spatial extent.
The investigation of the nonlocality can be rephrased as the inspection of the operator dependence, which was already discussed for the mesonic and baryonic systems \cite{Kawai:2017goq,Iritani:2018zbt,Aoki:2019gqt}.

Moreover, we have only calculated the DM cross section in the $SU(2)$ YMT, and the cases for $N_c \ge 3$ were just qualitative extrapolations using $1/N_c$ expansion in the present paper.
Since the interglueball cross section, being an $O(N_c^{-4})$ quantity, receives a correction of $O(N_c^{-6})$, we expect to complete the analysis of the $0^{++}$ glueball of $SU(N_c)$ YMT as the DM candidate for all $N_c$'s with the accuracy of $O(1\%)$ by accomplishing the calculations up to $N_c=10$.
This project is in our view not impossibly challenging.

\begin{acknowledgments}
This work was supported by ``Joint Usage/Research Center for Interdisciplinary Large-scale Information Infrastructures'' (JHPCN) in Japan (Project ID: jh180058-NAH).
The calculations were carried out on SX-ACE at RCNP/CMC of Osaka University.
MW was supported by the National Research Foundation of Korea (NRF) grant funded by the Korea government (MSIT) (No.~2018R1A5A1025563).
\end{acknowledgments}

\appendix

\section{\label{sec:glueballEFT}Derivation of Yukawa potential from glueball effective Lagrangian}

\begin{figure}[htbp]
  \begin{center}
    \begin{tabular}{c}
      \begin{minipage}{0.2\hsize}
        \begin{center}
          \includegraphics[clip, width=3cm]{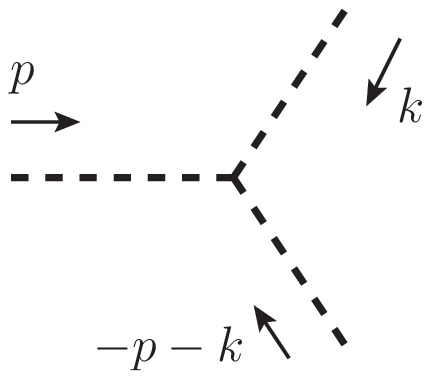}
        \end{center}
      \end{minipage}
      \begin{minipage}{0.8\hsize}
        \begin{center}
\begin{eqnarray}
&=&
\frac{i}{2f_\phi}
\Bigl[ 
p^2 + k^2 + (p + k )^2 
\Bigr]
+32i \frac{H_0}{f_\phi^3}
\nonumber
\end{eqnarray}
          \hspace{1.6cm}
        \end{center}
      \end{minipage}

    \end{tabular}
    \caption{Feynman rule for the glueball trilinear interaction.
    }
    \label{fig:glueball_EFT_feynman_rule}
  \end{center}
\end{figure}

We derive the nonrelativistic Yukawa potential from the glueball effective Lagrangian of Eq. (\ref{eq:glueballEFT}).
The Feynman rule of the trilinear interaction is given in Fig. \ref{fig:glueball_EFT_feynman_rule}.

We then calculate the t-channel one-glueball exchange process (see Fig. \ref{fig:Yukawa_potential}) with the trilinear interaction of Fig. \ref{fig:glueball_EFT_feynman_rule}:
\begin{eqnarray}
i{\cal M}
&=&
\frac{169}{4}
\frac{m_\phi^4}{f_\phi^2}
\frac{-i}{k^2 -m_\phi^2}
+i\frac{27}{4}
\frac{m_\phi^2}{f_\phi^2}
-\frac{i}{4 f_\phi^2} k^2
,
\label{eq:oneglueballexchange}
\end{eqnarray}
where $k$ is the exchanged momentum, while the momentum squared of the asymptotic glueballs are replaced by $m_\phi^2$, since we take the nonrelativistic approximation.
Here we used the relation $m_\phi = \sqrt{-4H_0 /f_\phi^2}$ to erase $H_0$.
The first term of the last equality of Eq. (\ref{eq:oneglueballexchange}) yields the Yukawa potential.
The fitting form of Eq. (\ref{eq:Yukawafit}) can be matched with the above one-glueball exchange amplitude as
\begin{equation}
V_Y^{(1)}
=
-
\frac{169}{16}
\frac{m_\phi^2}{f_\phi^2}
= 
\frac{169}{4}
\frac{H_0}{f_\phi^4}
.
\end{equation}
Here the additional factor $4m_\phi^2 = (\sqrt{2 m_\phi})^4$ in the denominator is coming from the normalization of nonrelativistic glueball state.

\bibliographystyle{Science}

\bibliography{full_paper_glueball_DM}

\end{document}